\documentclass[12pt]{article}

\usepackage[centertags]{amsmath}
\usepackage{amsfonts}
\usepackage{amssymb}
\usepackage{amsthm}
\usepackage{newlfont}
\usepackage{epsfig}
\usepackage{amscd}

\newcommand{\NN}{{\mathbb N}}

\newcommand{\RR}{{\mathbb R}}

\newcommand{\beq}{\begin{equation}}
\newcommand{\eeq}{\end{equation}}
\newcommand{\ba}{\begin{array}}
\newcommand{\ea}{\end{array}}
\newcommand{\bea}{\begin{eqnarray}}
\newcommand{\eea}{\end{eqnarray}}

\begin{document}

\begin{center}
{\large \sc \bf The Cauchy Problem on the Plane for the} 

\vskip 5pt
{\large \sc \bf Dispersionless Kadomtsev - Petviashvili Equation} 

\vskip 15pt

{\large  S. V. Manakov$^{1,\S}$ and P. M. Santini$^{2,\S}$}

\vskip 8pt

{\it 
$^1$ Landau Institute for Theoretical Physics, Moscow, Russia

\smallskip

$^2$ Dipartimento di Fisica, Universit\`a di Roma "La Sapienza" and\\
Istituto Nazionale di Fisica Nucleare, Sezione di Roma 1\\
Piazz.le Aldo Moro 2, I-00185 Roma, Italy}

\vskip 5pt

$^{\S}$e-mail:  {\tt manakov@itp.ac.ru, paolo.santini@roma1.infn.it}

\vskip 5pt

{\today}

\end{center}

\begin{abstract}
We construct the formal solution of the Cauchy problem for  
the dispersionless Kadomtsev - Petviashvili equation as application  
of the Inverse Scattering Transform for the vector field corresponding to 
a Newtonian particle in a time-dependent potential. This is in full analogy with the 
Cauchy problem for the Kadomtsev - Petviashvili equation, associated with the 
Inverse Scattering Transform of the time dependent Schr\"odinger operator for a 
quantum particle in a time-dependent potential.

\end{abstract}
\noindent
1.   
Dispersionless (or quasi-classical) limits of integrable partial differential equations (PDEs) arise in various 
problems of Mathematical Physics and   
are intensively studied in the recent literature (see, f.i., \cite{Kri,TT,DMT,K-MA-R,G-M-MA}). In particular, 
a quasi-classical dressing has been developed \cite{K-MA-R} 
for the prototypical example of the dispersionless Kadomtsev - Petviashvili (dKP) 
(or Khokhlov-Zabolotskaya) equation:
\beq
\label{dKP1}
u_{tx}+u_{yy}+(uu_x)_x=0,~~~
u=u(x,y,t)\in\RR,~~~~~x,y,t\in\RR.
\eeq 
      
In this paper we construct the formal solution of the Cauchy problem on the plane 
for the following  system of PDEs in 2+1 dimensions:
\beq
\label{dKP-system}
\ba{l}
u_{xt}+u_{yy}=-(uu_x)_x-v_xu_{xy}+v_yu_{xx}, ~~~~u,v\in\RR,~~x,y,t\in\RR,\\
v_{xt}+v_{yy}=-uv_{xx}-v_xv_{xy}+v_yv_{xx}
\ea
\eeq
and for its $v=0$ reduction, the dKP equation (\ref{dKP1}),   
as application of the recently developed   
Inverse Scattering Transform (IST) for vector fields \cite{MS}. Indeed the system (\ref{dKP-system}) 
arises as the compatibility condition of the Lax pair
\beq
\label{Laxpair1}
\hat L_1\psi=0,~~\hat L_2\psi=0,
\eeq
implying $[\hat L_1,\hat L_2]=0$, where $\hat L_1,~\hat L_2$ are the following vector fields:
\beq
\label{L1L2-syst}
\ba{l}
\hat L_1\equiv \partial_y+(p+v_x)\partial_x-u_x\partial_p, \\
\hat L_2\equiv \partial_t+(p^2+pv_x+u-v_y)\partial_x+(-pu_x+u_y)\partial_p.
\ea
\eeq
Setting $v=0$ in (\ref{L1L2-syst}), one obtains the Lax pair of the dKP equation, which  
was derived in \cite{DMT} taking the quasi-classical limit of the well-known Lax pair of the KP equation \cite{Dryuma,ZS}. 

We remark that, in the dKP reduction $v=0$, the two vector fields are Hamiltonian and the Lax pair (\ref{L1L2-syst}) 
takes the form   
\beq
\label{Laxpair2}
\ba{l}
\psi_y+p\psi_x-u_x\psi_p=\psi_y+\{H_1,\psi \}_{(p,x)}=0, \\
\psi_t+(p^2+u)\psi_x+(-pu_x+u_y)\psi_p=\psi_t+\{H_2,\psi \}_{(p,x)}=0,
\ea
\eeq 
in terms of the two Hamiltonians \cite{DMT}
\beq
\label{Hamiltonians}
H_1=\frac{p^2}{2}+u,~~~~
H_2=\frac{p^3}{3}+pu-\partial^{-1}_xu_y,
\eeq
where $\{\cdot,\cdot\}_{(p,x)}$ is the standard Poisson bracket with respect to the canonical variables $(p,x)$:
\beq
\label{Poisson}
\{f,g\}_{(p,x)}\equiv f_pg_x-f_xg_p,
\eeq
leading to the Hamiltonian form of dKP: ${H_1}_t-{H_2}_y+\{H_2,H_1\}_{(p,x)}=0$.

Since the Lax pair (\ref{Laxpair1}) of the dKP-like system (\ref{dKP-system}) is made of vector fields, 
Hamiltonian in the dKP reduction 
(\ref{dKP1}), the eigenfunctions satisfy the following basic properties.

\noindent
1) {\it The space of eigenfunctions is a ring}.  If $f_1,~f_2$ are two solutions of the Lax pair 
(\ref{Laxpair1}), then an arbitrary differentiable function $F(f_1,f_2)$ of them is a solution of (\ref{Laxpair1}).   

\noindent
2) {\it In the dKP reduction $v=0$, the space of eigenfunctions is also a Lie algebra, whose Lie bracket is 
the natural Poisson bracket (\ref{Poisson})}. If $f_1,~f_2$ are two solutions of the Lax pair (\ref{Laxpair2}), 
then their Poisson bracket $\{f_1,f_2\}_{(p,x)}$ is also a solution of (\ref{Laxpair2}).

\vskip 5pt\noindent
2. Now we consider the Cauchy problem for the dKP system (\ref{dKP-system}) 
and for the dKP equation (\ref{dKP1})     
within the class of rapidly decreasing real potentials $u,v$:
\beq
\label{localization}
u,v\to~0,~~(x^2+y^2)\to\infty, ~~u\in\RR,~~~(x,y)\in\RR^2,~~t>0,
\eeq
interpreting $t$ as time and the other two variables $x,y$ as space variables.  
To solve such a Cauchy problem by the IST method \cite{ZMNP}, we construct the 
IST for the operator $\hat L_1$, within the class of rapidly decreasing real potentials, 
interpreting the operator $\hat L_2$ as the time operator. 

The localization (\ref{localization}) of the 
potentials $u,~v$ implies that, if $f$ is a solution of $\hat L_1 f=0$, then 
\beq
\label{asymptf}
\ba{l}
f(x,y,p)\to f_{\pm}(\xi,p),\;\;y\to\pm\infty, \\
\xi:=x-py;
\ea
\eeq
i.e., asymptotically, $f$ is an arbitrary function of $\xi=x-py$ and $p$.

A central role in the theory is played by the two real Jost eigenfunctions $\varphi_{1,2}(x,y,p)$, the solutions of 
$\hat L_1\varphi_{1,2}=0$ uniquely defined by the asymptotics 
\beq
\label{asympt-varphi}
\varphi_{1}(x,y,p)\to \xi,~~~~\varphi_{2}(x,y,p)\to p, \;\;\;\;y\to -\infty.
\eeq
In this paper we often use the compact vector notation: $\vec f=(f_1,f_2)^T$. Then:
\beq
\label{def-vec-varphi}
\vec\varphi(x,y,p)\equiv \left(
\ba{l}
\varphi_1(x,y,p) \\
\varphi_2(x,y,p)
\ea
\right) \to \left(
\ba{l}
\xi \\
p
\ea
\right)\equiv \vec \xi,~~y\to -\infty.
\eeq 
The Jost eigenfunction $\vec\varphi$ is the solution of the linear integral equations 
$\vec\varphi=\vec \xi+\hat G(-v_x\vec\varphi_x+u_{x}\vec\varphi_p)$, for the Green's function $G(x,y,p)=\theta(y)\delta(x-py)$.

The $y=+\infty$ limit of $\vec\varphi$ defines the natural scattering vector $\vec\sigma$ for $\hat L_1$:
\beq
\label{def-S}
\displaystyle\lim_{y\to +\infty}\vec\varphi(x,y,p) \equiv 
\vec{\cal S}(\vec \xi)=\vec \xi+\vec\sigma(\vec \xi).
\eeq

The direct problem is the transformation from the real potentials $u,v$, functions of the two real variables $(x,y)$, 
to the two real scattering data $\sigma_1,\sigma_2$, the components of the scattering vector $\vec\sigma$, 
functions of the two real 
variables $(\xi,p)$. Therefore the mapping is consistent. The impact of the dKP reduction $v=0$ on these and other data 
will be shown below.  

A crucial role in the IST theory for the vector field $\hat L_1$  
is also played by the analytic eigenfunctions $\vec\psi_{\pm}(x,y,p)$, the solutions of 
$\hat L_1\vec\psi_{\pm}=\vec 0$  satisfying the integral equations 
\beq
\label{def-psi}
\ba{l}
\vec\psi_{\pm}(x,y,p)=
\int_{\RR^2}dx'dy'G_{\pm}(x-x',y-y',p)[-v_{x'}(x',y')\vec{\psi_{\pm}}_{x'}(x',y',p)+\\
u_{x'}(x',y')\vec{\psi_{\pm}}_p(x',y',p)]+\vec \xi,
\ea
\eeq
where $G_{\pm}$ are the analytic Green's functions
\beq
\label{Green_analytic}
G_{\pm}(x,y,p)=\pm\frac{1}{2\pi i[x-(p \pm i\epsilon) y]}.
\eeq
The analyticity properties of $G_{\pm}(x,y,p)$ in the complex $p$ - plane 
imply that $\vec\psi_{+}(x,y,p)$ and $\vec\psi_{-}(x,y,p)$ are 
analytic, respectively, in the upper and lower halves of the $p$ - plane, with 
the following asymptotics, for large $p$:
\beq
\label{asympt-psi}
\ba{l}
\vec\psi_{\pm}(x,y,p)=\vec \xi+\frac{1}{p}\vec U(x,y)+\vec O\left(\frac{1}{p^2}\right),~~|p|>>1, \\
\vec U(x,y)\equiv \left(
\ba{c}
-yu(x,y)-v(x,y) \\
u(x,y)
\ea
\right).
\ea
\eeq

It is important to remark that the analytic Green's functions (\ref{Green_analytic}) exhibit the following 
asymptotics for $y\to\pm\infty$:
\beq
\ba{l}
G_{\pm}(x-x',y-y',p)\to\pm\frac{1}{2\pi i[\xi-\xi'\mp i\epsilon]},\;\;y\to +\infty, \\
G_{\pm}(x-x',y-y',p)\to\pm\frac{1}{2\pi i[\xi-\xi'\pm i\epsilon]},\;\;y\to -\infty,
\ea
\eeq
entailing that {\it the $y=+\infty$ asymptotics of $\vec\psi_{+}$ and $\vec\psi_{-}$ are analytic 
respectively in the lower and upper halves of the complex plane $\xi$, while the $y=-\infty$ 
asymptotics of $\vec\psi_{+}$ and $\vec\psi_{-}$ are analytic respectively in the upper and lower 
halves of the complex plane $\xi$} (similar features have been observed first in \cite{MZ} and later in \cite{MS}).

The Jost eigenfunctions $\varphi_{1,2}$ form a basis; thus any solution $f$ of $\hat L_1f=0$ is a function of 
$\vec\varphi$. The analytic eigenfunctions $\vec\psi_{\pm}$ possess the representations:
\beq
\label{varphi-psi}
\vec\psi_{\pm}=\vec{\cal K}_{\pm}(\vec\varphi)=\vec\varphi+\vec\chi_{\pm}(\vec\varphi),
\eeq
defining the spectral data $\vec\chi_{\pm}$. 

Since the $y\to -\infty$ limit of (\ref{varphi-psi}) reads:
\beq
\label{lim-varphi-psi}
\displaystyle\lim_{y\to -\infty}\vec\psi_{\pm}-\vec \xi=\vec\chi_{\pm}(\vec \xi),
\eeq
the above analyticity properties of the LHS of (\ref{lim-varphi-psi}) in the complex 
$\xi$ - plane imply that $\vec\chi_{+}(\vec \xi)$ and $\vec\chi_{-}(\vec \xi)$ are analytic respectively in the upper and lower 
halves  of the complex plane $\xi$, decaying at $\xi\sim\infty$ like $O(\xi^{-1})$. Therefore their Fourier transforms 
$\tilde{\vec\chi}_{+}(\vec\omega)$ and $\tilde{\vec\chi}_{-}(\vec\omega)$ have support respectively on the positive and 
negative $\omega_1$ semi-axes. 

The spectral vectors $\vec\chi_{\pm}$ can be constructed from the scattering vector $\vec\sigma$  
through the following linear integral equations 
\beq
\label{Fourier-varphi-psi}
\ba{l}
\tilde{\vec\chi}_+(\vec\omega)+\theta(\omega_1)\left(\tilde{\vec\sigma}(\vec\omega)+
\int_{\RR^2}d\vec\eta ~\tilde{\vec\chi}_+(\vec\eta)Q(\vec\eta,\vec\omega)\right)=\vec 0,  \\
\tilde{\vec\chi}_-(\vec\omega)+\theta(-\omega_1)\left(\tilde{\vec\sigma}(\vec\omega)+
\int_{\RR^2}d\vec\eta ~\tilde{\vec\chi}_-(\vec\eta)Q(\vec\eta,\vec\omega)\right)=\vec 0,
\ea
\eeq 
involving the Fourier transforms $\tilde{\vec\sigma}$ and $\tilde{\vec\chi}_{\pm}$ of $\vec\sigma$ and ${\vec\chi}_{\pm}$:
\beq
\label{Fourier-sigma}
\tilde{\vec\sigma}(\vec\omega)=\int_{\RR^2}d\vec \xi\vec\sigma(\vec \xi)e^{-i\vec\omega\cdot\vec \xi},~~~
\tilde{\vec\chi}_{\pm}(\vec\omega)=\int_{\RR^2}d\vec \xi{\vec\chi}_{\pm}(\vec \xi)e^{-i\vec\omega\cdot\vec \xi}
\eeq
and the kernel:
\beq
\label{def-Q}
Q(\vec\eta,\vec\omega)=\int_{\RR^2}\frac{d\vec \xi}{(2\pi)^2}e^{i(\vec\eta-\vec\omega)\cdot\vec \xi}
[e^{i\vec\eta\cdot\vec\sigma(\vec \xi)}-1].
\eeq 
To prove this result, one first evaluates (\ref{varphi-psi}) at $y=+\infty$, obtaining 
\beq
\label{+lim-varphi-psi}
\left(\displaystyle\lim_{y\to \infty}\vec\psi_{\pm}-\vec \xi\right)=\vec\sigma(\vec \xi)+
\vec\chi_{\pm}(\vec \xi+\vec\sigma(\vec \xi)).
\eeq 
Applying the integral operator $\int_{\RR^2}d\vec \xi e^{-i\vec\omega\cdot\vec \xi}\cdot$ 
for $\omega_1>0$ and $\omega_1<0$ respectively to equations (\ref{+lim-varphi-psi})$_{+}$ and 
(\ref{+lim-varphi-psi})$_{-}$, using the above analyticity properties  
and the Fourier representations of ${\vec\chi}_{\pm}$ and $\vec\sigma$, one obtains equations 
(\ref{Fourier-varphi-psi}).

The reality of the potentials: $u,v\in\RR$ implies that, for $p \in\RR$,  
$\overline{\vec\varphi}=\vec\varphi$, $\overline{\vec\psi}_+=\vec\psi_-$; consequently: 
$\overline{\vec\sigma}=\vec\sigma$, $\overline{\vec\chi}_+=\vec\chi_-$.

\vskip 5pt
\noindent
3. An inverse problem can be constructed from equations (\ref{varphi-psi}). 
Subtracting $\vec \xi$ from equations (\ref{varphi-psi})$_{-}$ and (\ref{varphi-psi})$_{+}$,  
applying respectively the analyticity projectors $\hat P_{+}$ and $\hat P_{-}$: 
\beq
\hat P_{\pm}\equiv \pm\frac{1}{2\pi i}\int_{\RR}\frac{dp'}{p'-(p\pm i\epsilon)}. 
\eeq
and adding up the resulting equations, one obtains the following 
nonlinear integral equation for the Jost eigenfunction $\vec\varphi$:
\beq
\label{varphi-int-equ2}
\ba{l}
\vec\varphi(x,y,p)+\frac{1}{2\pi i}\int_{\RR}\frac{dp'}{p'-(p+i\epsilon)}\vec\chi_-(\vec\varphi(x,y,p')) - \\
\frac{1}{2\pi i}\int_{\RR}\frac{dp'}{p'-(p-i\epsilon)}\vec\chi_+(\vec\varphi(x,y,p'))=\vec\xi.
\ea
\eeq
Once $\vec\varphi$ is reconstructed from (\ref{varphi-int-equ2}), the analytic eigenfunctions follow from (\ref{varphi-psi}), 
and $u,v$ from equation (\ref{asympt-psi}). This inversion procedure was first introduced in \cite{Manakov1} and also used in 
\cite{MS}.

\vskip 5pt
\noindent
4. As $u,v$ evolve in time according to (\ref{dKP-system}), the $t$-dependence of the  
spectral data $\vec{\cal S}$ and $\vec{\cal K}_{\pm}$,  
defined in (\ref{def-S}) and (\ref{varphi-psi}), is described by the equations:
\beq
\label{t-dep-Sigma}
\ba{l}
\Sigma_1(\xi,p,t)=t\left(\Sigma_2(\xi-p^2t,p,0)\right)^2+\Sigma_1(\xi-p^2t,p,0), \\
\Sigma_2(\xi,p,t)=\Sigma_2(\xi-p^2t,p,0),
\ea
\eeq
where $\Sigma_1$ and $\Sigma_2$ are the two components of the vector $\vec\Sigma$, identifiable with each of 
the spectral vectors $\vec {\cal S}$ and $\vec{\cal K}_{\pm}$. To prove it, we first observe that 
\beq
\label{def-phi1}
\ba{l}
\phi_1(x,y,t,p)\equiv \varphi_1(x,y,t,p)-t\varphi^2_2(x,y,t,p), \\
\phi_2(x,y,t,p)\equiv \varphi_2(x,y,t,p)
\ea
\eeq
are a basis of   
common Jost eigenfunctions of $\hat L_1$ and $\hat L_2$. The $y=+\infty$ limit of equation $\hat L_2\phi_2=0$ 
yields ${{\cal S}_2}_t+p^2{{\cal S}_2}_{\xi}=0$, while the $y=+\infty$ limit of equation $\hat L_2\phi_1=0$ 
yields $(\partial_t+p^2\partial_{\xi})({\cal S}_1-t{\cal S}^2_2)=0$, whose solutions are (\ref{t-dep-Sigma}) 
for $\vec {\cal S}$. Analogously, 
\beq
\ba{l}
{\pi_{\pm}}_1(x,y,t,p)\equiv {\psi_{\pm}}_1(x,y,t,p)-t{\psi_{\pm}}^2_2(x,y,t,p), \\
{\pi_{\pm}}_2(x,y,t,p)\equiv {\psi_{\pm}}_2(x,y,t,p)
\ea
\eeq
are a basis of common analytic eigenfunctions of $\hat L_1$ and $\hat L_2$; therefore 
\beq
{\pi_{\pm}}_1={{\check{\cal K}}_{\pm 1}}(\phi_1,\phi_2),~~~{\pi_{\pm}}_2={\check{\cal K}_{\pm 2}}(\phi_1,\phi_2), 
\eeq
for some functions ${\check{\cal K}_{\pm 1,2}}$ depending on $x,y,t,p$ only through $\vec\phi$.  
Comparing at $t=0$ these equations 
with equations (\ref{varphi-psi}), one expresses ${\check{\cal K}_{\pm {1,2}}}$  
in terms of ${{\cal K}_{\pm}}_{1,2}$, obtaining equations (\ref{t-dep-Sigma}) for 
${{\cal K}_{\pm}}_{1,2}$.

We observe the unusual resonant character of the explicit $t$-dependence (\ref{t-dep-Sigma}) of the spectral data, 
if compared to the more elementary one, obtained in \cite{MS}, for the heavenly equation \cite{Pleb}.

\vskip 5pt
\noindent
5. In the Hamiltonian dKP reduction $v=0$, the transformations $\vec\xi\to\vec{\cal S}(\vec\xi)$, 
$\vec\xi\to\vec{\cal K}_{\pm}(\vec\xi)$ are constrained to be canonical:
\beq
\label{data-constraint}
\{{\cal S}_1,{\cal S}_2\}_{\vec\xi}=\{{{\cal K}_{\pm}}_1,{{\cal K}_{\pm}}_2\}_{\vec\xi}=1.
\eeq
To prove it, we observe that the Poisson bracket of the eigenfunctions $\varphi_1$ and $\varphi_2$ is also an eigenfunction:
$\varphi_3\equiv \{\varphi_1,\varphi_2\}_{(x,p)},~~\hat L_1\varphi_3=0$. 
Using the asymptotics (\ref{asympt-varphi}), one infers that $\varphi_3\to 1,~$ at $y\to -\infty$; therefore, by 
uniqueness, $\varphi_3=1$. Evaluating now the Poisson bracket $\varphi_3$ at $y=+\infty$ and using  
(\ref{def-S}), one obtains the constraint (\ref{data-constraint}) for $\vec{\cal S}$. We also observe that the 
eigenfunctions $\{{\psi_+}_1,{\psi_+}_2\}_{(x,p)}$ and 
$\{{\psi_-}_1,{\psi_-}_2\}_{(x,p)}$ are analytic in the upper and lower $p$ plane and go to $1$ at $|p|\to\infty$. 
Since $1$ is also an eigenfunction, by uniqueness they are identically $1$: $\{{\psi_{\pm}}_1,{\psi_{\pm}}_2 \}_{(x,p)}=1$. 
Therefore, from the equations:
\beq
\{{\psi_{\pm}}_1,{\psi_{\pm}}_2 \}_{(x,p)}=
\{{{\cal K}_{\pm}}_1,{{\cal K}_{\pm}}_2 \}_{(\varphi_1,\varphi_2)}\{\varphi_1,\varphi_2 \}_{(x,p)}=1,
\eeq
consequence of (\ref{varphi-psi}), one infers  the constraints (\ref{data-constraint}) for $\vec{\cal K}_{\pm}$.

\vskip 5pt
\noindent
6. It is well-known (see, f.i., \cite{CH}) that linear first order PDEs 
like (\ref{Laxpair1}),(\ref{L1L2-syst}) are intimately related to systems of ordinary differential equations describing their 
characteristic curves. The Hamiltonian dynamical systems associated with the vector fields $\hat L_{1,2}$ 
of dKP are: 
\beq
\label{flow1}
\ba{ll}

\hat L_1:   & \left\{
\ba{l}
\frac{dx}{dy}=p=\{H_1,x \}_{(p,x)}, \\
\frac{dp}{dy}=-u_x=\{H_1,p \}_{(p,x)}
\ea
\right.
\ea\eeq
\beq
\label{flow2}
\ba{ll}
\hat L_2:  & \left\{
\ba{l}
\frac{dx}{dt}=p^2+u,=\{H_2,x \}_{(p,x)}, \\
\frac{dp}{dt}=-pu_x+u_y=\{H_2,p \}_{(p,x)},
\ea
\right.
\ea
\eeq
Therefore {\it the dKP equation  characterizes the class of time - dependent  
potentials for wich the Newtonian flow (\ref{flow1}) commutes with a flow with cubic, in the momentum $p$, 
Hamiltonian}. 
 
There is also a deep connection between the above IST and the $y$-scattering theory for the commuting flows 
(\ref{flow1}) and (\ref{flow2}). Let $\vec\phi(x,y,t,p)$ be the basis of common eigenfunctions of $\hat L_1$ and 
$\hat L_2$ defined in (\ref{def-phi1}); then, solving the system 
$\vec\omega=\vec\phi(x,y,t,p)$ with respect to $x$ and $p$ (assuming local invertibility), one obtains the following 
common solution of (\ref{flow1}) and (\ref{flow2}):
\beq
\vec\omega=\vec\phi(x,y,t,p)~~\Leftrightarrow~~\left(
\ba{c}
x \\
p
\ea
\right)=\vec r(y,t;\vec\omega)~\sim~\left(
\ba{c}
\omega_2y+\omega^2_2t+\omega_1 \\
\omega_2 
\ea
\right),~~y\sim -\infty.
\eeq
The $y=+\infty$ limit of the solution $\vec r(y,t;\vec\omega)$:
\beq
\left(
\ba{c}
x \\
p
\ea
\right)~\sim~\left(
\ba{c}
\Omega_2(\vec\omega)y+\Omega^2_2(\vec\omega)t+\Omega_1(\vec\omega) \\
\Omega_2(\vec\omega) 
\ea
\right),~~y\sim +\infty
\eeq
defines the scattering vector $\vec\Delta(\vec\omega)=\vec\Omega(\vec\omega)-\vec\omega$ 
of (\ref{flow1}) and (\ref{flow2}), which is connected to the IST data $\vec{\cal S}$ by inverting the 
system $\vec\omega=\vec{\cal S}(x-py-p^2t,p,0)$ with 
respect to $x$ and $p$:
\beq
\vec\omega=\vec{\cal S}(x-py-p^2t,p,0)~\Leftrightarrow~\left(
\ba{c}
x \\
p
\ea
\right)=\left(
\ba{c}
\Omega_2(\vec\omega)y+\Omega^2_2(\vec\omega)t+\Omega_1(\vec\omega) \\
\Omega_2(\vec\omega) 
\ea
\right).
\eeq
The transformation $\vec\omega \to \vec\Omega(\vec\omega)$ is clearly canonical: 
$\{{\vec\Omega}_1,{\vec\Omega}_2\}_{(\omega_1,\omega_2)}=1$.

Since the dynamical system (\ref{flow1}) describes the motion of a Newtonian 
particle in the plane subjected to a generic time - dependent potential $u(x,y)$, as a byproduct of the IST of 
this paper one can reconstruct, from the scattering 
vector $\vec\Delta(\vec\omega)$ of the dynamical system (\ref{flow1}), the time dependent potential $u$.

\vskip 5pt
\noindent
{\it Remark 1}. The are two other ways to do the inverse problem. The first one is the linear version of the nonlinear problem  
(\ref{varphi-int-equ2}), obtained {\it exponentiating the Jost and analytic eigenfunctions} used so far. Consider the 
following scalar functions:
\beq
\Phi(x,y,p;\vec\alpha)\equiv e^{i\vec\alpha\cdot\vec\varphi(x,y,p)},~~~
\Psi_{\pm}(x,y,p;\vec\alpha)\equiv e^{i\vec\alpha\cdot\vec\psi_{\pm}(x,y,p)},~~~~\vec\alpha\in\RR^2.
\eeq
Due to the ring property of the space of eigenfunctions, also $\Phi(x,y,p;\vec\alpha)$ and $\Psi_{\pm}(x,y,p;\vec\alpha)$ 
are eigenfunctions; $\Phi(x,y,p;\vec\alpha)$ is characterized by the asymptotics 
$\Phi\to exp(i\vec\alpha\cdot\vec \xi),~y\to -\infty$, while $\Psi_{\pm}(x,y,p;\vec\alpha)$ are analytic respectively in the 
upper and lower halves of the $p$ plane, with asymptotics: $\Psi_{\pm}=exp(i\vec\alpha\cdot\vec \xi)
[1+p^{-1}\vec\alpha\cdot\vec U(x,y)+O(p^{-2})]$.

Exponentiating the representations (\ref{varphi-psi}), one obtains the expansions of the analytic eigenfunctions 
$\Psi_{\pm}$ in terms of the Jost eigenfunction $\Phi$:
\beq
\label{Phi-Psi}
\ba{l}
\Psi_{\pm}(x,y,p;\vec\alpha)=\Phi(x,y,p;\vec\alpha)+
\int_{\RR^2}d\vec\beta K_{\pm}(\vec\alpha,\vec\beta)\Phi(x,y,p;\vec\beta), \\
K_{\pm}(\vec\alpha,\vec\beta)\equiv \int_{\RR^2}\frac{d\vec \xi}{(2\pi)^2}e^{i(\vec\alpha-\vec\beta)\cdot\vec \xi}
[e^{i\vec\alpha\cdot\vec\chi_{\pm}(\vec \xi)}-1].
\ea
\eeq
Multiplying the equations (\ref{Phi-Psi})$_{+}$ and (\ref{Phi-Psi})$_{-}$ by 
$exp(-i\vec\alpha\cdot\vec \xi)$, subtracting $1$, applying respectively $\hat P_{-}$ and $\hat P_{+}$,  
and adding the resulting equations, one obtains the following {\it linear integral equation} for $\Phi$: 
\beq
\label{Phi-int-equ}
\ba{l}
\Phi(p;\vec\alpha)+\frac{1}{2\pi i}\int_{\RR}\frac{dp'}{p'-(p+i\epsilon)}

\int_{\RR^2}d\vec\beta K_-(\vec\alpha,\vec\beta)\Phi(p';\vec\beta)e^{i\vec\alpha\cdot(\vec \xi(p)-\vec \xi(p'))} - \\
~~ \\
-\frac{1}{2\pi i}\int_{\RR}\frac{dp'}{p'-(p-i\epsilon)}
\int_{\RR^2}d\vec\beta K_+(\vec\alpha,\vec\beta)\Phi(p';\vec\beta)e^{i\vec\alpha\cdot(\vec \xi(p)-\vec \xi(p'))} =
e^{i\vec\alpha\cdot\vec\xi(p)}, 
\ea
\eeq
in which we have omitted, for simplicity, the parametric dependence on $(x,y)$. 
Once $\Phi$ is reconstructed from (\ref{Phi-int-equ}) and, via (\ref{Phi-Psi}), $\Psi_{\pm}$ are also known, 
the potentials  are reconstructed in the usual way from the asymptotics of $\Psi_{\pm}$. 

The third version of the inverse problem is a more traditional (nonlinear) Riemann-Hilbert (RH) problem.  
Solving the algebraic system (\ref{varphi-psi})$_-$ with respect to $\vec\varphi$: $\vec\varphi=L(\vec\psi_{-})$ 
(assuming local invertibility) and replacing   
this expression in the algebraic system (\ref{varphi-psi})$_+$, one obtains the representation of the analytic eigenfunction 
$\vec\psi_{+}$ in terms of the analytic eigenfunction $\vec\psi_{-}$:
\beq
\label{RH}
\vec\psi_{+}=\vec{\cal R}(\vec\psi_{-})=\vec\psi_{-}+\vec R(\vec\psi_{-}),~~p\in\RR,
\eeq
which defines a {\it vector nonlinear RH problem on the real $p$ axis}. The RH 
data $\vec{\cal R}$ are therefore constructed from the data $\vec{\cal K}$ by algebraic manipulation. Viceversa, 
given the RH data $\vec R$, one constructs the solutions $\vec\psi_{\pm}$ of the nonlinear RH problem (\ref{RH}) and, 
via the asymptotics (\ref{asympt-psi}), the potentials.  

As for the other spectral data, one can show that 
the $t$-dependence of $\vec{\cal R}$ is described by (\ref{t-dep-Sigma}) and the dKP constraint reads 
$\{{\cal R}_1,{\cal R}_2\}_{(\xi,p)}=1$, while the reality constraint takes the form: 
$\vec{\cal R}(\overline{ \vec{\cal R}(\bar{\vec\xi},\lambda)   },\lambda)=\vec\xi,~\forall\vec\xi$, 
for $p \in\RR$. 

\vskip 2pt
\noindent
{\it Remark 2}. Dressing schemes can be formulated from the three different inverse problems 
presented in this paper in a straightforward way. 

\vskip 2pt
\noindent
{\it Remark 3}. The IST constructed in this paper allows one to 
solve the Cauchy problem for the whole hierarchy of PDEs arising from the commutativity equation $[\hat L_1,\hat L^{(n)}_2]=0$, 
where the coefficients of the vector field $\hat L^{(n)}_2$ are polynomials in $p$ of arbitrary degree $n\in\NN$.

\vskip 2pt
\noindent
{\it Remark 4}. There are deep similarities between the Cauchy problem for dKP and the Cauchy problem for the heavenly equation, 
 recently solved in \cite{MS}, since they are both based on the IST for Hamiltonian vector fields (the dKP equation is 
actually a geometric reduction of the heavenly equation \cite{DMT}). There is, however, 
an important difference between these two cases. The vector fields of the dKP equation  
contain partial derivatives with respect to the spectral parameter $p$,   
unlike the case of the heavenly equation \cite{MS}. 
 
\vskip 10pt
\noindent
{\bf Acknowledgements}. The visit of SVM to Rome was supported by the INFN grant 2005 and by the RFBR grant 04-01-00508.  
PMS thanks M. Dunajski for pointing out reference \cite{DMT}.

\end{document}